# Three-dimensional honeycomb carbon: Junction line distortion and novel emergent fermions

Junping Hu[a,b,1], Weikang Wu[a,1], Chengyong Zhong[c], Ning Liu[b], Chuying Ouyang[d], Hui Ying Yang[a*], Shengyuan A. Yang[a*]

Graphical abstract

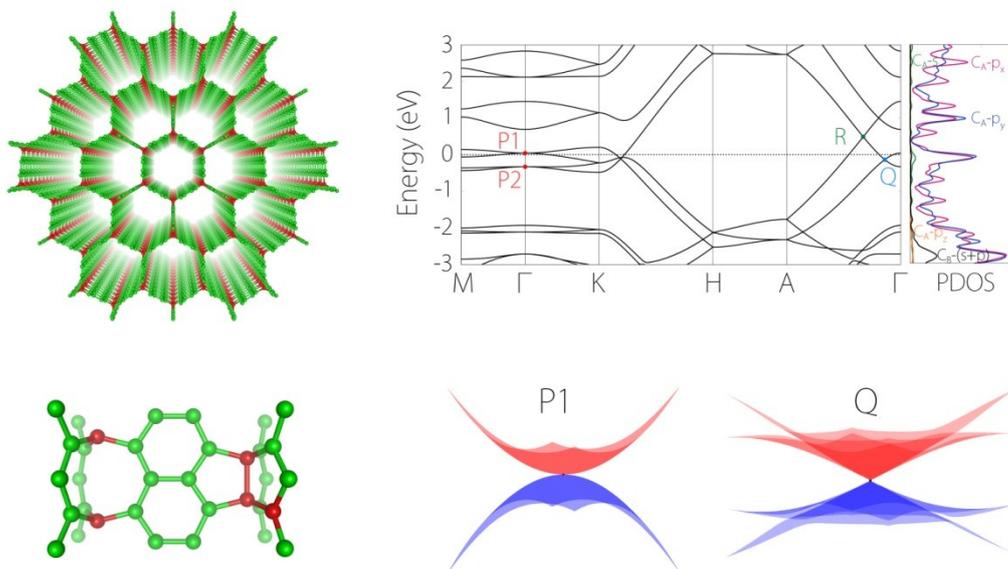

# Three-dimensional honeycomb carbon: Junction line distortion and novel emergent fermions

Junping Hu[a,b,1], Weikang Wu[a,1], Chengyong Zhong[c], Ning Liu[b], Chuying Ouyang[d], Hui Ying Yang[a*], Shengyuan A. Yang[a*]

a. Research Laboratory for Quantum Materials, Singapore University of Technology and Design, Singapore 487372, Singapore

b. School of Science, Nanchang Institute of Technology, Nanchang 330099, China

c. Institute for Quantum Information and Spintronics, School of Science, Chongqing University of Posts and Telecommunications, Chongqing 400065, China.

d. Department of Physics, Laboratory of Computational Materials Physics, Jiangxi Normal University, Nanchang 330022, China.

These authors contributed equally to this work.
*Corresponding Author:
yanghuiying@sutd.edu.sg
shengyuan_yang@sutd.edu.sg;

**ABSTRACT:** Carbon enjoys a vast number of allotropic forms, each possessing unique properties determined by the lattice structures and bonding characters. Here, based on first-principles calculations, we propose a new three-dimensional carbon allotrope—hC28. We show that hC28 possesses exceptional energetic, dynamical, thermal, and mechanical stability. It is energetically more stable than most other synthesized or proposed carbon allotropes. The material has a relatively small bulk modulus, but is thermally stable at temperatures as high as 2000 K. The structural, mechanical, x-ray diffraction, and electronic properties are systematically investigated. Particularly, we show that its low-energy band structure hosts multiple unconventional emergent fermions, including the quadratic-contact-point fermions, the birefringent Dirac fermions, and the triple-point fermions. We construct effective models to characterize each kind of fermions. Our work not only discovers a new carbon allotropic form, it also reveals remarkable mechanical and electronic properties for this new material, which may pave the way towards both fundamental studies as well as practical applications.

## 1. Introduction

Carbon, one of the most abundant elements in the universe, has never ceased to amaze and surprise us. Due to its ability to form versatile *sp* hybridized bonds, carbon can generate a vast number of allotropic forms[1-2]. The most well-known are diamond with *sp*$^3$ bonding, graphite (and graphene) with *sp*$^2$ bonding, as well as fullerenes with mixed bonding characters[3]. The different structures and bonding characters can lead to dramatically different properties among these carbon materials. For example, while diamond is one of the hardest materials, graphite can be easily exfoliated and hence used as lubricants. As another example, graphene[4]—a single layer of carbon atoms—possesses extraordinary electronic mobility[5] and thermal conductivity[6], whilst diamond is a good insulator. Thus, the carbon materials provide a clear demonstration of the structure-activity relationship[7]. This also indicates that determining the precise microstructure of carbon allotropes is very critical.

The first-principles density functional theory (DFT) calculations have been proved to be a powerful tool in determining the microscopic atomic structures. Although advanced experimental techniques such as scanning tunneling microscopy (STM), X-ray/electron diffraction, and tunneling electron microscopy (TEM) can directly image the microscopic structures, each has its limitations and often it is difficult to resolve small structural distortions in practice, especially if the sample is not of a single crystalline phase. For such cases, DFT calculations will play a crucial role in clarifying the correct ground state structure. Furthermore, DFT studies can predict new structures, particularly when combined with structural search algorithms[8-10]. Many new carbon allotropes have been predicted in the past[11-22]. Some of them have been successfully verified, or related to previously unknown phases in experiment[12,22].

Meanwhile, carbon allotropes have been shown to provide a unique platform for realizing novel topological band structures. In so-called topological metals or semimetals, novel quasiparticles emerge at protected band crossing points, giving rise to condensed matter realization of many exotic fermions, such as Weyl fermions[23-24], Dirac fermions[25-27], nodal-loop/nodal-chain fermions[28-32], nodal-surface fermions[33-35], and etc. While a majority of works in the field have been focused on materials involving heavy elements, expecting that nontrivial band topology could be driven by the strong spin-orbit coupling (SOC), the carbon materials offer a distinct alternative. Because of the negligible SOC strength, spin can be considered as a dummy degree of freedom for carbon materials (unless magnetism appears), and the emergent fermions can be regarded as "spinless" or "spin-orbit-free" fermions, which are fundamentally distinct from the "spinful" fermions in spin-orbit-coupled systems[36]. Indeed, several examples have already been proposed, such as the 2D Dirac fermions in graphene[5,37], the 3D Weyl fermions and the nodal-loop fermions in nanostructured carbons[29,38-39], the triple-point fermions and Hopf-link fermions in 3D pentagon carbon[21], and the nodal-surface fermions in graphene networks[33]. Owing to the large number of carbon allotrope structures with different crystalline symmetries, one expects that more exotic fermions are waiting to be explored in carbon.

In this work, we propose a new three-dimensional (3D) carbon allotrope, termed as honeycomb-carbon-28 (denoted as hC28), which has exceptional stability and hosts novel spin-orbit-free emergent fermions. This study is also motivated by a recent experiment which synthesized a new carbon structure by using carbon sublimation method[40]. In that study, a 3D honeycomb carbon model was proposed as the candidate structure for the synthesized material[40]. As shown in Figure 1(b), this structure (denoted as S1 structure) is built from carbon atoms exclusively with $sp^2$ bonding, and it can be viewed as patches of graphene nanoribbons (running along *c*-axis) connected at junction lines, forming a super honeycomb structure in top view. Subsequent experimental and theoretical studies have revealed many interesting properties of this material[41-49]. However, the microscopic structure for the material has not been fully clarified. Later, in a brief comment[50], it was suggested that after relaxation, structure S1 will transform into the structure shown in Figure 1(c) (denoted as S2 structure)[12], in which bonds form between the C atoms on the junction lines, such that these atoms tend to become *sp³* hybridized. Our proposed hC28 structure, shown in Figure 1(d), is closely related to the structures S1 and S2. Focusing on the junction line atoms, S2 can be viewed as a result of dimerization of these atoms as compared with S1, and the junction lines dimerize in a symmetric manner. In comparison, in hC28, the dimerization occurs in an asymmetric pattern between neighboring junction lines. We show that hC28 is energetically more stable than structures S1 and S2, and in fact better than most other carbon allotropes. By investigating the phonon spectrum, we find that structure S1 is unstable and both S2 and hC28 are dynamically stable. We further perform *ab-initio* molecular dynamics (AIMD) simulations and find that S1 will transform into hC28 at room temperature, whereas hC28 can be stable even at 2000 K. We systematically study the structural, mechanical, and x-ray diffraction properties for hC28, and compare them with other carbon allotropes. In addition, we show that the electronic band structure for hC28 features several novel spin-orbit-free emergent fermions, including the quadratic-contact-point fermions, birefringent Dirac fermions, as well as triple-point fermions. They will lead to interesting unconventional features in physical properties, such as magnetic response or magneto-transport. Given the exceptional stability of hC28, it is very likely that it is the true structure for the material synthesized in Ref[40]. Our work not only offers a highly promising candidate structure for experiment, it also reveals the extraordinary properties of this new structure, which may find great potential in future applications.

## 2. Computational methods

Our first-principles calculations are based on the density functional theory (DFT) using the plane-wave pseudopotentials[51-52] as implemented in the Vienna *ab initio* Simulation Package (VASP)[53-54]. The exchange-correlation functional is modeled in the generalized gradient approximation (GGA) with the Perdew-Burke-Ernzerhof (PBE)[55] realization. Carbon $2s^2 2p^2$ electrons are treated as valence electrons in all the calculations. A cutoff energy of 550 eV is employed for the plane wave expansion of the wave functions. The Brillouin zone is sampled with $3 \times 3 \times 5$ Monkhorst-Pack $k$-point mesh[56] for the structural optimization, and with $7 \times 7 \times 15$ mesh for the electronic structure calculations. The convergence criteria for the total energy and ionic forces are set to be $10^{-8}$ eV and $10^{-6}$ eV/Å, respectively. Phonon spectra calculations are performed using the Phonopy package[57]. In our AIMD simulations, the NVT canonical sampling is performed by integrating the equations of motion at 0.5 fs time intervals, and the temperature is controlled via a Nosé-Hoover thermostat[58-59]. At each time step, the total energy is evaluated to an accuracy of $10^{-5}$ eV/atom using a plane-wave energy cutoff of 500 eV.

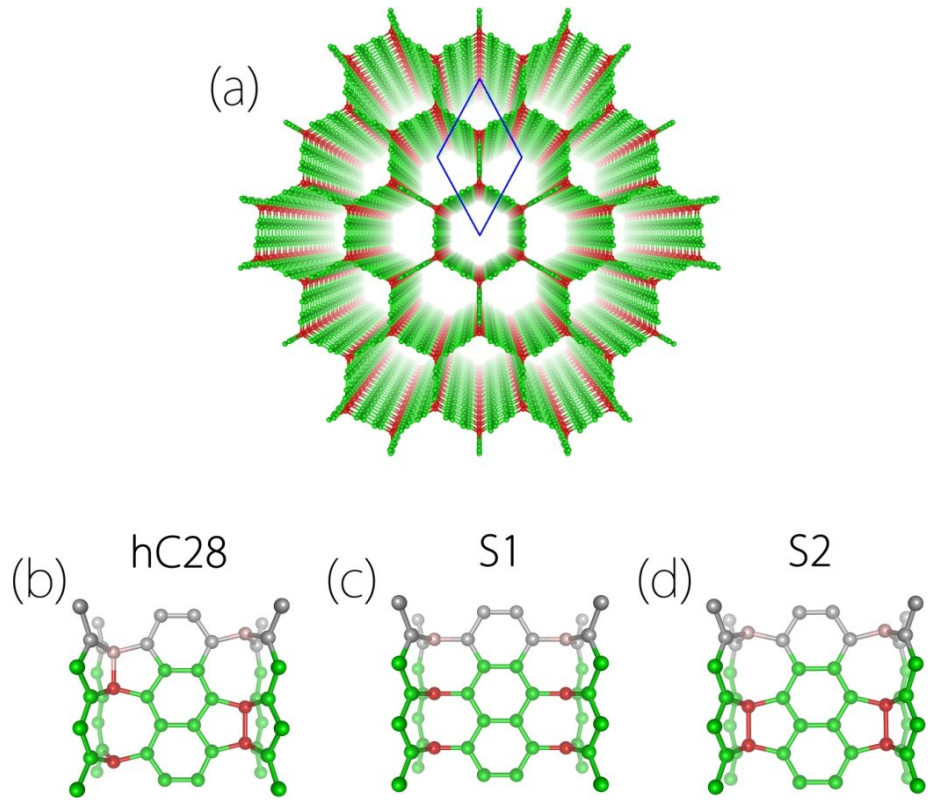

Figure 1 Lattice structure of 3D honeycomb carbon allotrope. (a) Top perspective view of 3D carbon honeycomb. (b-d) Side views of the unit cell for (b) hC28, (c) S1, and (d) S2 structures. The unit cell contains 28 carbon atoms. The red and green atoms represent the carbon atoms located in the junction lines and graphene nanoribbons, respectively. The gray colored atoms represent the atoms in the neighboring unit cell.

## 3. Results and discussion

### 3.1 Lattice structure

The lattice structure for hC28 is schematically shown in Figure 1(b). For comparison, in Figure 1(c, d), we also show the closely related structures S1 and S2. For hC28 and S2, a unit cell contains 28 carbon atoms, whereas for S1 structure, a unit cell contains

14 atoms. Here, for better comparison, we double the unit cell for S1 and plot the $1 \times 1 \times 2$ supercell in Figure 1(c).

All three structures can be viewed as consisting of zigzag graphene nanoribbons connected to form a honeycomb lattice in top view [see Figure 1(a)]. Three neighboring nanoribbons are jointed at a junction line. The key difference between the three structures lies in the configuration of the junction-line atoms. In S1, the junction line atoms are evenly distributed along the $c$-axis, with a separation of 2.434 Å between two neighboring atoms, as indicated in Figure 2(b). This interatomic distance is much larger than the typical C-C bond length (~1.54 Å), so there is no covalent bonding between them, and the junction-line atoms each forms $sp^2$ bonding with its three neighbors in the $x$-$y$ plane. Thus, S1 corresponds to an all-$sp^2$ bonding network. In contrast, for hC28 and S2, the junction-line configuration is distorted. If we focus only on the junction-line atoms, the distortion corresponds to the dimerization of a 1D chain of carbon atoms [see Figure 1(b, d) and Figure 2(a, c)]. This dimerization leads to a doubled period along $c$ for hC28 and S2 as compared to S1. For hC28, the shorter (longer) neighbor separation distance is 1.605 Å (3.273 Å), while the corresponding values for S2 are 1.600 Å and 3.189 Å. Hence, two closer neighbors form a C-C bond. For S2, the dimerization pattern is the same for all junction lines [see Figure 1(d)], whereas for hC28, the pattern is asymmetric between any two neighboring lines [see Figure 1(b)].

The hC28 structure belongs to the space group No. 194 ($D_{6h}$-4), whereas both S1 and S2 belong to the space group No. 191 ($D_{6h}$-1). Comparing the symmetry operations for S2 and hC28, the key difference is that the two-fold rotation $C_{2z}$ for S2 is replaced by the two-fold screw rotation $S_{2z} = \{C_{2z}|00\frac{1}{2}\}$ for hC28. Focusing on the hC28 structure, its lattice parameters are: $a = b = 10.032$ Å, and $c = 4.878$ Å. The volume of its unit cell is fairly large, of about 425.19 Å$^3$. There are four irreducible atomic Wyckoff positions: 6$h$ (0.4631, -0.0737, 0.2500), 6$h$ (0.5468, 0.0936, 0.2500), 12$k$ (0.4180, 0.8361, 0.4901), and 4$f$ (0.3333, 0.6667, 0.4145). The corresponding Wyckoff positions for the other two structures as well as other structural parameters are listed in Table S-1 of the Supplementary Information.

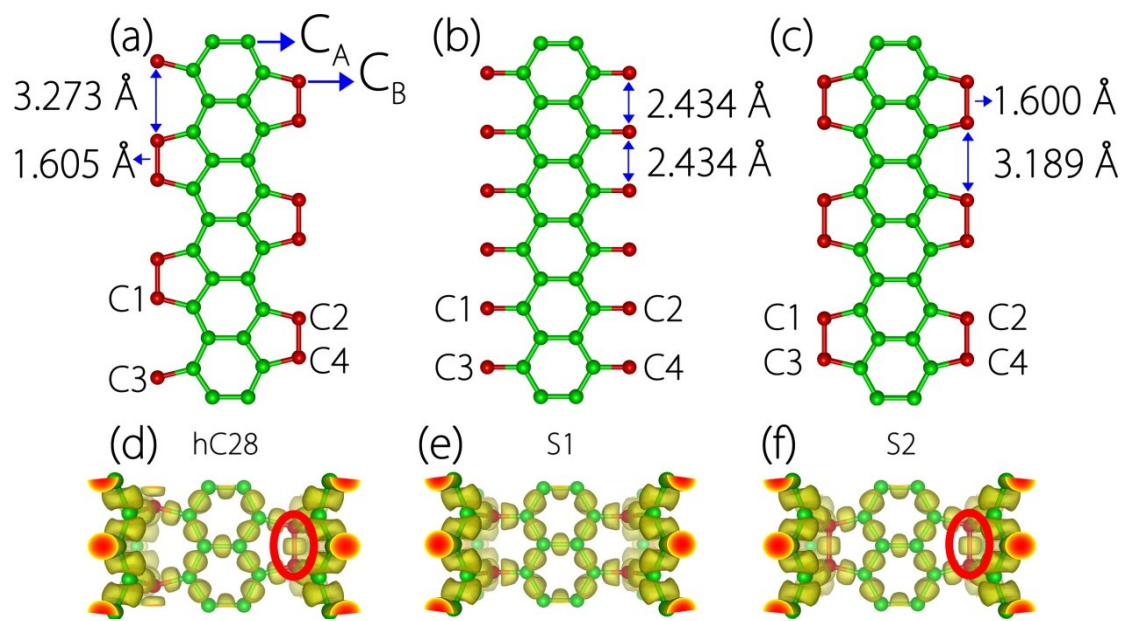

Figure 2 Configuration for the armchair graphene nanoribbons between two junction lines in (a) hC28, (b) S1, and (c) S2, respectively. Here, C1, C2, C3, C4 represent the four inequivalent carbon atoms on the junction lines. Symbols C$_A$ and C$_B$ denote the two kinds of carbon atoms in hC28 with $sp^2$ and $sp^3$ hybridization character, respectively. (d-e) Maps of electron localization function (ELF) isosurfaces for (d) hC28, (e) S1, and (f) S2, computed for a value of 0.698. The circles in (d) and (f) highlight the bonding features between two junction-line atoms.

### 3.2 Energetic and dynamical stability

The proposed hC28 structure has a lower total energy as compared to S1 and S2. In terms of total energy for the unit cell in Figure 1 (b-d), hC28 is 3.724 eV lower than

S1 and 0.840 eV lower than S2. This demonstrates that hC28 is energetically the most stable one among the three candidate structures.

Figure 3 shows the calculated total energy versus volume per atom for hC28 compared to other carbon phases, including S1[40], S2[12], oC16[60], 3D-C5[21], bco-C16[61], graphite, M-carbon[62], and diamond. Among them, diamond is in an all-$sp^3$ bonding network, hC28, S2 and oC16 carbon are in a mixed $sp^2$-$sp^3$ bonding network, and the other carbon structures are all-$sp^2$ bonding networks. As shown in Figure 3, while hC28 is higher in energy than diamond and graphite, it is energetically more favorable than all other carbon phases examined here. This demonstrates the exceptional stability of the proposed structure. The equilibrium volume for hC28 is larger than most other carbon phases, and slightly smaller than S1 and S2, indicating its lower atomic density. By fitting the calculated total energy versus volume curve to Murnaghan's equation of state, we obtain a bulk modulus of 155 GPa for hC28, which is much smaller than the values for diamond (~450 GPa) and graphite (~280 GPa), consistent with its lower atomic density. The key equilibrium structural parameters for hC28 and other carbon phases are listed in Table 1 for comparison.

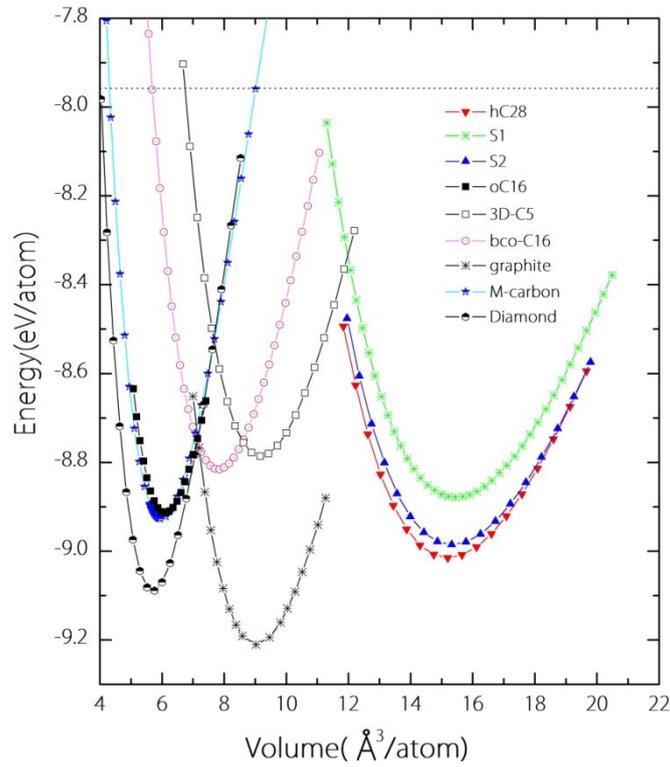

Figure 3 Calculated total energies versus volume per atom for hC28 compared with several other carbon allotropes. The dashed line represents the energy level of carbyne chain.

---

Next, we investigate the dynamical stability of the structure, which can be inferred from the phonon spectrum. Figure 4 shows the phonon spectra of the three structures hC28, S1, and S2. Obviously, a branch with huge imaginary frequency can be observed for S1 in Figure 4(b), indicating that the structure is dynamically unstable. As we show below, the S1 structure will transform into the hC28 structure at finite temperature, and the transformation is associated with the distortion of the junction line configuration. In comparison, the phonon spectra for both hC28 and S2 show no imaginary frequency modes [see Figure 4 (a, c)]. Thus, these two structures are dynamically stable. The highest phonon frequencies for hC28 and S2 are 1591 cm$^{-1}$ and 1613 cm$^{-1}$, respectively, which are close to that of graphite (~1610 cm$^{-1}$[63]).

Table 1 Optimized structural parameters (space group, lattice parameters $a$, $b$, and $c$, volume $V_0$, bond lengths $d_{C-C}$), total energy $E_{tot}$, and bulk modulus $B_v$ for hC28 compared with other carbon allotropes.

| Structure | Method | a (Å) | b (Å) | c (Å) | $V_0$ (Å$^3$/atom) | $d_{C-C}$ (Å) | $E_{tot}$ (eV/atom) | $B_v$ (Gpa) |
|---|---|---|---|---|---|---|---|---|
| hC28 (P6$_3$/mmc) | PBE | 10.032 | | 4.878 | 15.19 | 1.41~1.52 | -9.018 | 155 |
| S1 (P6/mmm) | PBE | 10.125 | | 4.869 | 15.43 | 1.41~1.49 | -8.885 | 153 |
| S2 (P6/mmm) | PBE | 10.094 | | 4.859 | 15.31 | 1.40~1.52 | -8.988 | 152 |
| oC16 (Cmmm) | PBE | 14.981 | 2.585 | 2.525 | 6.11 | 1.35~1.59 | -8.914 | 510 |
| 3D-C5 (I4$_1$/amd) | PBE | 7.146 | | | 9.16 | 1.37~1.44 | -8.786 | 257 |
| bco-C$_{16}$ (Imma) | PBE | 7.856 | 4.908 | 3.258 | 7.85 | 1.39~1.47 | -8.816 | 314 |
| Graphite P6$_3$/mmc) | PBE | 2.470 | | 6.827 | 9.02 | 1.43 | -9.210 | 279 |
| | Exp[64] | 2.460 | | 6.704 | 8.78 | 1.42 | | 286 |
| M-carbon (C2/m) | PBE | 9.089 | 2.496 | 4.104 | 5.95 | 1.49~1.61 | -8.927 | 420 |
| Diamond (Fd$\bar{3}$m) | PBE | 3.586 | | | 5.76 | 1.55 | -9.089 | 452 |
| | Exp[64] | 3.567 | | | 5.67 | 1.54 | | 446 |

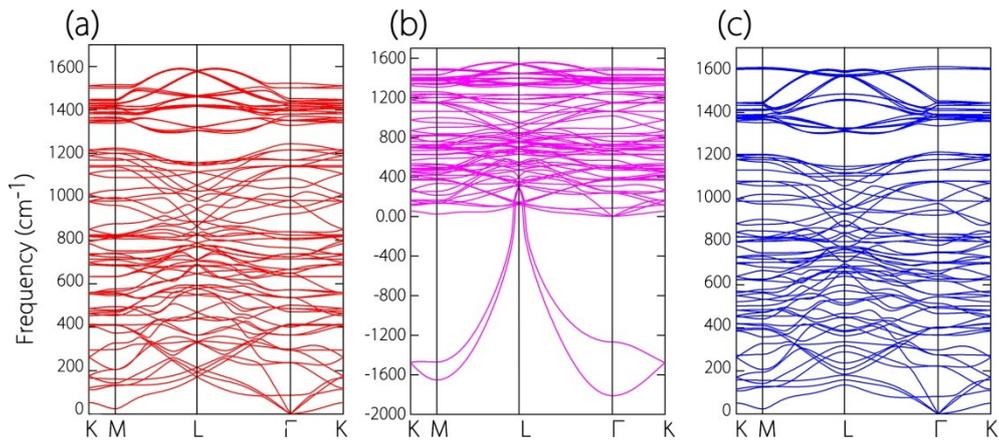

Figure 4 Calculated phonon spectra for the structures (a) hC28, (b) S1, and (c) S2.

## 3.3 Thermal stability and transition energy barrier

To further examine the thermal stability of hC28, we perform the AIMD simulations at 300 K. The simulations for other two structures S1 and S2 are also done for comparison. As illustrated in Figure S-1, we observe that hC28 retains its integrity throughout the simulation period, showing that it is stable at room temperature. In fact, we have also performed the simulation at elevated temperatures up to 2000 K and found that hC28 remains to be stable, indicating that hC28 indeed has excellent stability.

For the other two structures, we find that S2 is also stable at 300 K, but S1 is not. A closer look shows that under thermal fluctuation, the junction-line atoms in S1 would soon re-arrange and dimerize and automatically transform into the hC28 structure.

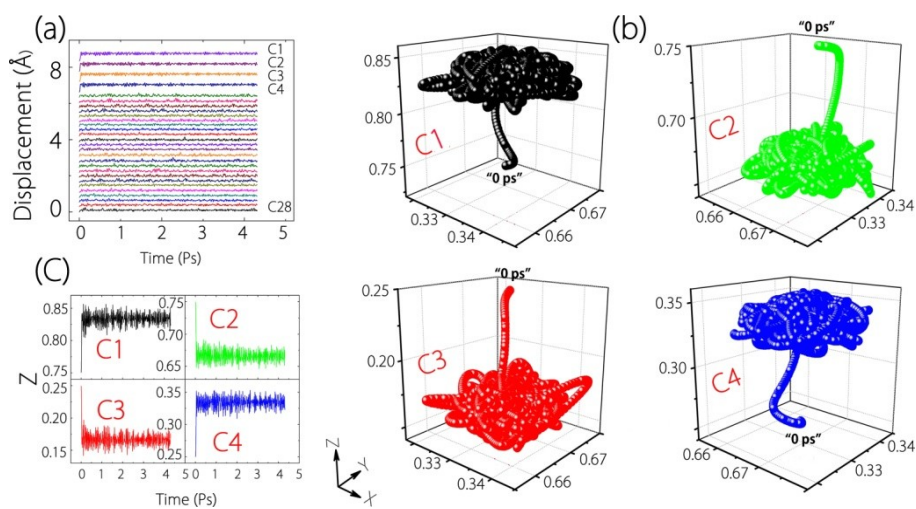

Figure 5 (a) Result from AIMD simulation, including the time-resolved displacement of all atoms in S1. (b) Evolution of the positions for the four atoms in 3D space. (c) Zoom in view for the displacement in z-direction of the four junction-line atoms (C1,

C2, C3, C4) in (a). The label of "0 ps" represents the initial location. The result shows that under thermal fluctuation, S1 transforms into the hC28 structure.

Since both hC28 and S2 are metastable and they are closely related structures, we also investigate the transition energy barrier between the two phases. It is evaluated by using the climbing image nudged elastic band (CI-NEB) technique[65-66]. As illustrated in Figure 6, the phase transition is mainly related to the re-arrangement of the carbon atoms on the junction lines. Here, two configurations for hC28 are chosen to be the initial and final states (see Figure 6), and S2 is chosen to be one of the transition states. The result shows that the transition energy barrier is fairly large ~2.1 eV per unit cell, much larger than the energy difference between hC28 and S2 (~0.87 eV). Thus, hC28 and S2 each is expected to be stable at room temperature, and there is unlikely to be automatic transformation (or resonance) between the two structures.

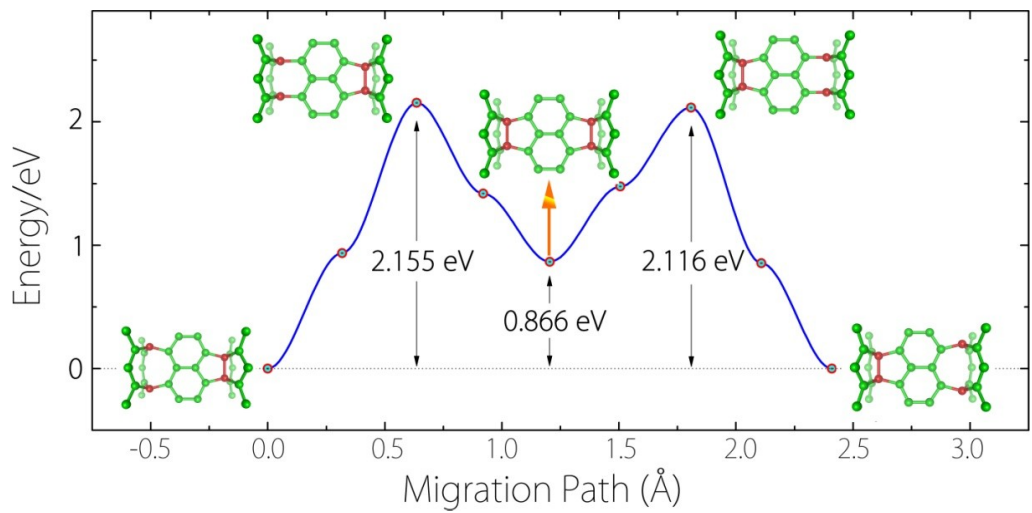

Figure 6 Evaluation for the phase transition barrier from hC28 to S2, and from S2 to hC28.

## 3.4 Elastic property, mechanical stability, and XRD spectra

We have performed a series of calculations to obtain the elastic constants of hC28 and to infer its mechanical stability. Due to the structural symmetry, there are five independent elastic constants $C_{11}$, $C_{12}$, $C_{13}$, $C_{33}$, and $C_{44}$. Their calculated values are listed in Table 2, along with results for S1, S2 and other typical materials. The values of the elastic constants meet the mechanical stability criteria for hexagonal phase[67], namely $C_{44} > 0$, $C_{11} > |C_{12}|$, and $(C_{11}+2\,C_{12})\,C_{33} > 2\,C_{13}^{2}$.

Table 2 Estimated elastic constants, bulk modulus ($B_v$), shear modulus ($G_v$), and Young's modulus for hC28, S1, and S2, compared with other materials. All data are in unit of GPa.

| Structure | $C_{11}$ | $C_{12}$ | $C_{13}$ | $C_{33}$ | $C_{44}$ | $B_v$ | $G_v$ | Young's modulus |
|---|---|---|---|---|---|---|---|---|
| hC28 | 157.5 | 147.9 | 47.7 | 591.4 | 103.5 | 155 | 86.6 | 218 |
| S1 | 174.2 | 164.3 | 39.4 | 541.4 | 85.2 | 153 | 78.2 | 200 |
| S2 | 163.3 | 153.8 | 37.8 | 584.0 | 37.8 | 152 | 61.5 | 162 |
| Steel | | | | | | 160 | 79.3 | 200 |
| Glass | | | | | | 35-55 | 26.2 | 50-90 |
| Diamond | | | | | | 443 | 478.0 | |
| Copper | | | | | | | 44.7 | 117 |

In Table 2, we also list the calculated bulk modulus ($B_v$), shear modulus ($G_v$), and Young's modulus. One observes that these mechanical properties for hC28 are quite similar to those of steel. This suggests that hC28 may replace steel or used in combination with steel in potential applications.

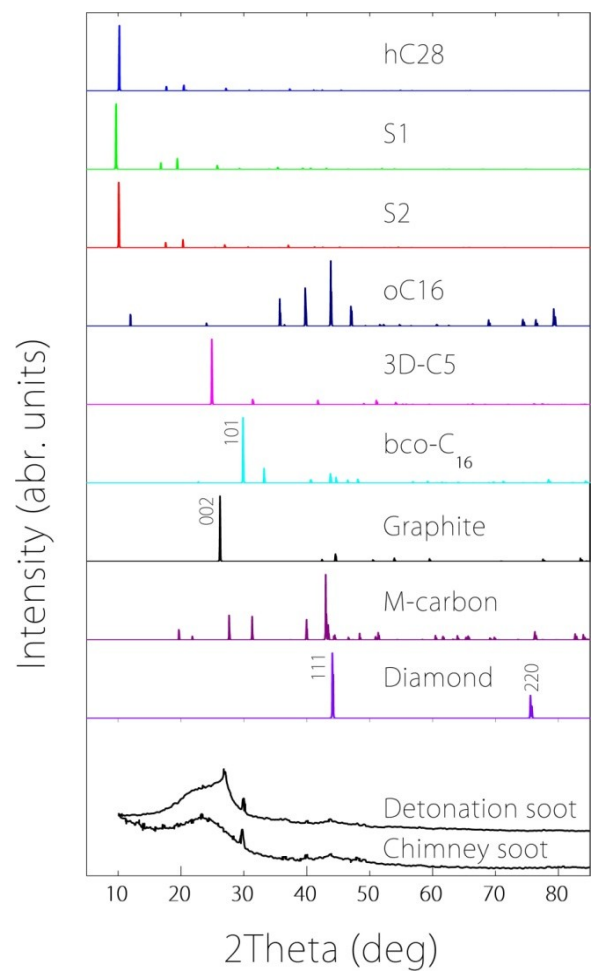

Figure 7 X-ray diffraction (XRD) patterns. Simulated XRD patterns for hC28, compared with other carbon phases, as well as experimental XRD patterns for detonation soot of TNT, diesel oil[68] and chimney soot[69]. Here, the x-ray wavelength is 1.5406 Å with a copper source.

---

To facilitate the experimental characterization of this new carbon phase, we simulate the x-ray diffraction (XRD) spectrum for hC28, and compare it to the other carbon phases, namely S1, S2, oC16, 3D-C5, bco-C16, graphite, M-carbon, diamond, and also to the experimental data from detonation soot of TNT and diesel oil[68] and chimney soot[69]. As we know, the measured XRD spectra reveal a substantial amount of amorphous carbon and provide clear evidence for several crystalline phases in the reconstructed specimen, such as graphite [(002) diffraction peak near 26.5°], bco-C16 carbon [(101) diffraction peak near 30°], and diamond [(111) diffraction peak near 43.7°][61]. The prominent peak for hC28 is located at 10.15°, which is close to that of S2 (~10.15°) and slightly larger than that of S1 (~9.65°). This peak can be discerned from peaks of other carbon phases, as shown in Figure 7. Thus, although XRD cannot easily differentiate hC28 from S2, it will be of great value to distinguish hC28 (or S2) from other carbon phases.

### 3.4 Electronic band structure and emergent fermions

In Figure 8, we show the electronic band structure for hC28 without SOC. The band structures with and without SOC exhibit little difference, due to the negligible SOC strength of carbon atoms. One observes that the material is a semimetal with several band-crossing points near the Fermi level. From the projected density of states (PDOS) in Figure 8, one notices that the bands around the Fermi level are dominated by the $C_A$-$p_x$ and $C_A$-$p_y$ orbitals, where $C_A$ refers to the carbon atoms with $sp^2$ bonding character [see Figure 2(a)]. Of the band-crossing points, we will pay special attention to the following four: P1 and P2 at the $\Gamma$ point, as well as Q and R on the $\Gamma$-$A$ path. We shall see that: points P1 and P2 are doubly-degenerate points with quadratic dispersions; Q is a four-fold degenerate birefringent Dirac point; and R is a triply-degenerate point. In the following, we analyze them one by one.

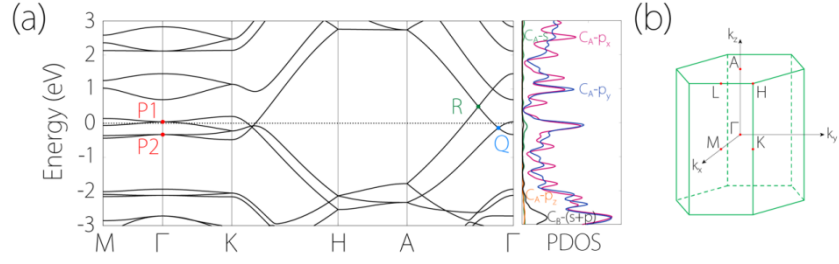

Figure 8 (a) Calculated electronic band structure for hC28 without SOC. The right panel shows the projected density of states (PDOS). The four crossing points in the low-energy band structure are labeled as P1, P2, Q, and R, respectively. In PDOS, $C_A$ and $C_B$ represent two kinds of carbon atoms with $sp^2$ and $sp^3$ hybridization, respectively. (b) Brillouin zone with high-symmetry points labeled.

### 3.4.1 Quadratic contact point

Let us first consider the two doubly-degenerate points P1 and P2 located at the $\Gamma$ point. Through a careful scan of the band structure around P1, we find that the energy dispersion of the two crossing bands is of quadratic-type in all three directions. In Figure 9, we plot the band dispersion around P1 in the $k_x$-$k_y$ plane, which clearly shows the quadratic band contact feature.

The nature of quadratic dispersion as well as the low-energy quasiparticles can be captured by constructing a $k \cdot p$ model. The symmetry at the $\Gamma$ point is of $D_{6h}$, and we find that the two degenerate states at P1 correspond to the two-dimensional irreducible representation $E_{1g}$. Using the two states as basis, one can construct the effective Hamiltonian constrained by the following symmetries: the sixfold rotation $C_{6z}$, the twofold rotation $C_{2x}$, the inversion $P$, and the time reversal $T$. Here, the space-time inversion $PT$ symmetry can be represented as $PT = \mathcal{K}$ with $\mathcal{K}$ the complex conjugation, which ensures that $H(\mathbf{k})$ is real. Then, up to quadratic order, we obtain the 2 × 2 effective Hamiltonian

$$H(\mathbf{k}) = \epsilon_0(\mathbf{k})\sigma_0 - 2\lambda k_x k_y \sigma_x + \lambda(k_x^2 - k_y^2)\sigma_z$$

where $\epsilon_0(\mathbf{k}) = C_0 + C_1(k_x^2 + k_y^2) + C_2 k_z^2$, and the $\sigma$'s are the Pauli matrices. The material-specific parameters $\lambda$ and $C_i$ can be determined by fitting the DFT band structure. From the effective model, we indeed see that the dispersion is of quadratic type around P1, with the eigenvalues $E(k_\perp) = \epsilon_0(k_\perp) \pm \lambda k_\perp^2$ in the $k_x$-$k_y$ plane ($k_\perp$ is the magnitude of a wave vector in the $k_x$-$k_y$ plane), and the two bands are degenerate along the $k_z$ axis.

The other point P2 about 0.5 eV below the Fermi level also shows a quadratic feature in the $k_x$-$k_y$ plane. The two degenerate states at P2 belong to the two-dimensional irreducible representation $E_{2u}$ of the $D_{6h}$ group. We find that the form of the effective model around point P2 is same as in Eq.(1), except for a sign change of the $\sigma_x$ term. The points P1 and P2 thus represent quadratic contact points, and the low-energy quasiparticles around them are the quadratic-contact-point fermions.

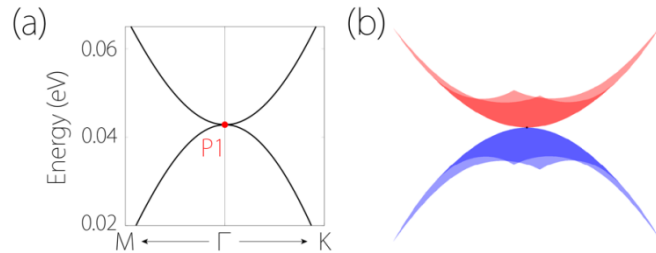

Figure 9 (a) Enlarged low-energy band structure of hC28 around point P1 along the $M$-$\Gamma$-$K$ path. (b) Band dispersion around point P1 in the $k_x$-$k_y$ plane, showing a quadratic dispersion.

### 3.4.2 Birefringent Dirac point

Next, let's consider the Dirac point Q located on the $\Gamma$-$A$ path as displayed in Figure 10. The Dirac point is formed by two crossing bands each is doubly degenerate, so the Dirac point is four-fold degenerate. The linear band crossing along $\Gamma$-$A$ is protected because the two crossing bands belong to two distinct 2D irreducible representations $E_1$ and $E_2$ of the $C_{6v}$ symmetry group for $\Gamma$-$A$. The Dirac point is an accidental crossing point, meaning that it is originated from the inverted band ordering of $E_1$ and $E_2$ bands between $\Gamma$ and $A$. Due to time reversal symmetry, there is a pair of such Dirac points on the two sides of $\Gamma$ point.

The dispersion around the Dirac point in the $k_x$-$k_y$ plane exhibits interesting features as shown in Figure 10(b). One observes that the four bands completely split in directions different from the $k_z$ axis. This is because deviating from the high-symmetry path $\Gamma$-$A$, there is no protection for the double degeneracy of the band and the degeneracy is generally lifted. Dirac point with such kind of dispersion is known as birefringent Dirac point[70-73], because there are two different Fermi velocities in the $k_x$-$k_y$ plane which may lead to birefringent refractions at p-n junction interfaces[73].

To characterize the emergent birefringent Dirac fermions, we construct the $k \cdot p$ effective model around Q, following the similar method mentioned above. Using the $E_1$ and $E_2$ states as basis, the effective $4 \times 4$ Hamiltonian around Q up to linear order takes the form

$$H(\boldsymbol{q}) = Cq_z + Bq_x\sigma_x \otimes \sigma_y + Bq_y\sigma_y \otimes \sigma_0 + Aq_z\sigma_z \otimes \sigma_0$$

Here the wave vector $q$ is measured from Q, the energy of the Dirac point is set as the zero energy, and the real parameter $A$, $B$, $C$ are the material specific parameters. The term $Ck_z$ represents a tilt of the spectrum along $k_z$. Such birefringent Dirac fermions have been studied in the context of cold atoms[70-72] and recently found in the CaAgBi family materials with strong SOC[73]. Our result here shows the first example of spin-orbit-free birefringent Dirac fermions.

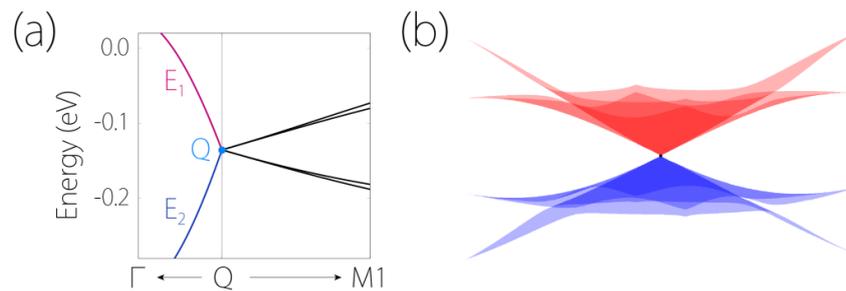

Figure 10 (a) Enlarged low-energy band structure for hC28 around point Q and (b) its band dispersion in the $k_x$-$k_y$ plane. The irreducible representations of bands on $\Gamma$-$A$ are indicated. The path $Q$-$M1$ is along the $k_x$ direction perpendicular to the $\Gamma$-$A$ path.

### 3.4.3 Triply degenerate point

For the band-crossing point R along the $\Gamma$-$A$ path, as shown in Figure 11(a), one notes that it is located at the intersection between a doubly degenerate band and a nondegenerate band, so it has a three-fold degeneracy. The crossing bands correspond to the $E_2$ and $B_1$ irreducible representations of the $C_{6v}$ symmetry group for $\Gamma$-$A$. Due to time reversal symmetry, there is a pair of such triply degenerate points on the two sides of $\Gamma$ on the $k_z$-axis. The band dispersion in the $k_x$-$k_y$ plane for a constant $k_z$ value around R is plotted in Figure 11(b-d). Exactly at R, one observes a Dirac-cone-like spectrum formed by the top and bottom bands, with the middle band crossing the intersection point, leading to the triple degeneracy. For the $k_x$-$k_y$ planes below or above R, two of the three bands still stick together forming a doubly degenerate point, while the remaining one is detached. One notes that since the ordering between the $E_2$ band and the $B_1$ band is inverted between $\Gamma$ and $A$, they must cross each other, forming the triply degenerate point.

Using the $E_2$ and $B_1$ states as basis, we can obtain the effective Hamiltonian (up to linear order) around the R point, given by

$$H(q) = Cq_z I_{3\times 3} + \begin{bmatrix} Aq_z & iBq_x & iBq_y \\ -iBq_x & 0 & 0 \\ -iBq_y & 0 & 0 \end{bmatrix}$$

Here the wave vector $q$ and the energy are measured from R, and the materials-specific parameters $A$, $B$ and $C$ can be determined by fitting the DFT result. This model characterizes the low-energy quasiparticle excitations around R, which are the triple-point fermions. Previous works have reported triple-point fermions in a few noncentrosymmetric spin-orbit-coupled materials[74-80], in contrast, their appearance here is in a spin-orbit-free system with centrosymmetry.

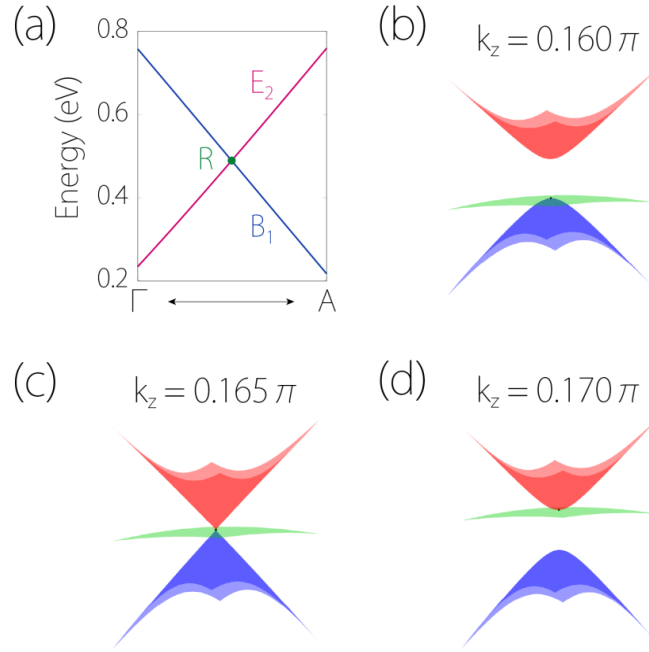

Figure 11 (a) Enlarged low-energy band structure for hC28 around point R. (b-d) The band dispersions in the $k_x$-$k_y$ plane for a constant $k_z$ (b) below, (c) at, and (d) above the R point. The irreducible representations of bands are indicated.

## 4. Conclusions

In this work, we have proposed a new carbon allotrope—hC28. It is related to two other honeycomb carbon structures S1 and S2, which had been postulated to explain an experimentally synthesized new carbon phase, but we find that hC28 is more stable than the other two. We show that hC28 possesses exceptional stability. (i) It is energetically more stable than most other carbon phases (except diamond and graphite) proposed or synthesized to date. (ii) It is dynamically stable without any soft phonon modes. (iii) It has excellent thermal stability, and can maintain its structure even at 2000 K. The other structure S1 would automatically transform to hC28 at room temperature. Based on these findings, it is quite plausible to suggest that hC28 may be the correct phase that was observed in experiment. In addition, we investigate the elastic properties of the material and demonstrate its mechanical stability. We find that the mechanical properties of hC28 are quite similar to steel, suggesting its potential to substitute or used in combination with steel in applications. Furthermore, we unveil interesting topological features in the electronic band structure of hC28. We show that there are multiple unconventional emergent fermions at low energy, including the quadratic-contact-point fermions, the birefringent Dirac fermions, and the triple-point fermions. We construct the effective $k \cdot p$ models to characterize each kind of fermions. It should be emphasized that these novel fermions are spin-orbit-free, which are fundamentally distinct from their counterparts studied in strongly spin-orbit-coupled materials. The spin-orbit-free quadratic-contact-point fermions and the birefringent Dirac fermions are uncovered here for the first time. Our work thus not only discovers a new carbon allotropic form, it also reveals remarkable mechanical and electronic properties for this new material, which may pave the way towards both fundamental studies as well as practical applications.

**Acknowledgments**

The authors thank S. Wu and D. L. Deng for valuable discussion. This work is supported by This research is supported by the National Research Foundation, Prime Minister's Office, Singapore under its NRF-ANR Joint Grant Call (NRF-ANR Award No. NRF2015-NRF-ANR000-CEENEMA), Natural Science Foundation of China (NSFC Grant No. 11564016), Natural Science Foundation of Jiangxi Province (No. 20171BAB216014), and Singapore Ministry of Education Academic Research Fund Tier 2 (MOE2015-T2-2-144). We acknowledge computational support from the Texas Advanced Computing Center and the National Supercomputing Centre Singapore.